# Figure-of-merit for Semi-transparent Solar Cells

**Arun Kumar*†, Sonia Rani†, Dhriti Sundar Ghosh***

Thin-film and Photovoltaics Laboratory, Department of Physics, Indian Institute of Technology Bhilai, Raipur, India.

**arunbagga@iitbhilai.ac.in*

**dhriti.ghosh@iitbhilai.ac.in*

*†Equal Contribution*

**Abstract:** Semi-transparent Solar Cells (ST-SCs) has emerged as one of the most prominent energy harvesting technology that combines the benefits of light transparency and light-to-electricity conversion. The biggest opportunities for such technologies lie in their integration as windows and skylights within energy-sustainable buildings or combining them with other solar cell technologies in tandem configuration. The performance of ST-SCs is mainly determined by the trade-off between the competing parameters of the capability to convert the incident light into electricity while allowing some parts to transmit providing transparency through the device. Depending on the target application, the selection of ST-SCs is a tricky affair as some devices might offer high efficiency but compromises transparency and vice-versa. On the other way around, this is again not helped by the fact that due to advancements in materials engineering, processing, and characterization, a vastly different combination of efficiency and transparency has been reported by research groups. So, in order to quantify the performance of ST-SCs, we proposed, a figure-of-merit (FoM) which can be used as a tool that can help in analysing and comparing the performance among various ST-SCs. The defined FoM focuses on the power conversion efficiency of the device, bifaciality factor, transmittance in the desired region, and that corresponding to 550 nm wavelength. Additionally, in this work, we have been shown how the proposed FoM can be correlated for tandem and building-integrated photovoltaics applications. Based on these resultant parameters, FoM is calculated and compared for different device architectures available in the literature. The proposed FoM shall serve as a meaningful guiding path to the researchers for the development of advanced ST-SCs.

## INTRODUCTION

Third-generation solar cells, in a very short time, have reached efficiencies that are comparable to the other solar technologies making them a robust technology for the future generation of solar cells. Apart from continuously improving efficiency in the last 5 years reaching the level of silicon technology, properties like mechanical flexibility, lightweightness, colorful, low-cost, customizable shapes, and semitransparency have made organic solar cells (OSCs) and perovskite solar cells (PvSCs), most exciting technologies as far as solar energy harvesting is concerned. Among these properties, being able to make them semi-transparent with still having respectable efficiency has made new avenues in terms of their usage and applications, especially in the building-integrated photovoltaics (BIPV) domain.

Although solar cells have already stepped out as an energy-efficient technology, it requires a lot of space for their installations. In the case of skyscraper buildings covering only a small area on the rooftop with solar cells will not supply a sufficient amount of energy. As the vertical area in such buildings is much larger compared to the rooftop area, another approach could be of using semi-transparent solar cells on façade or windows which absorb some part of the incoming light for electricity generation and transmit some portion to provide visibility inside the building during the daytime. By implementing such an approach, the entire building can be turned into a vertical power generator.

Generally, solar cells are designed to absorb the visible part concentrated in the narrow wavelength regime of 375-700nm of the solar spectrum reaching the earth crust (AM 1.5G) which accounts for ~42.3% of the total spectrum with peak intensity at about 550nm thereby making the device non-transparent. There are various impressive ways to make semi-transparent solar cells (ST-SCs). The easiest of all is to reduce the active layer thickness which can increase the transmittance of the device but the concern with this method lies in that due to less absorption of light, one has to compromise with efficiency. In addition, reduction of the thickness of the absorbing layer, also often leads to the problem related to the uniformity of the absorbing layer especially when deposited using solution processed[1]. Another approach is to tune the bandgap of the absorbing layer which can help balance the efficiency and semitransparency for ST-SCs. Bandgap tuning is more prominent in PvSCs and quantum dot solar cells owing to the ease of doing it by compositional engineering and size control of the quantum dots, respectively. In addition, ST-SCs are incomplete without a good top transparent contact. A ST-SCs make use of different types of top transparent electrodes (TEs) such as carbon nanotubes (CNTs), graphene layer, ultrathin metallic films (UTMFs), metallic nanowires, transparent conductive oxides (TCOs), and dielectric-metal-dielectric (DMD) structures [2–8]. Apart from this, index-matching capping layers such as Lithium Fluoride (LiF), Magnesium Flouride ($MgF_2$), etc. are also utilized for light management purposes to enhance the transparency and as well as efficiency by the selective transmission of wavelength through the device and by optical interference phenomenon within the stack respectively [9,10]. It also has to be noted that the use of transparent electrodes at both ends also contributes towards the reduction of efficiency compared to a cell incorporating a thick metallic electrode due to the loss of reflected photons from the back electrode. In a nutshell, these two parameters i.e., semitransparency and efficiency are always in a race against each other, and decoupling or better trade-off between these two parameters holds the key to realizing efficient ST-SCs with good transparency.

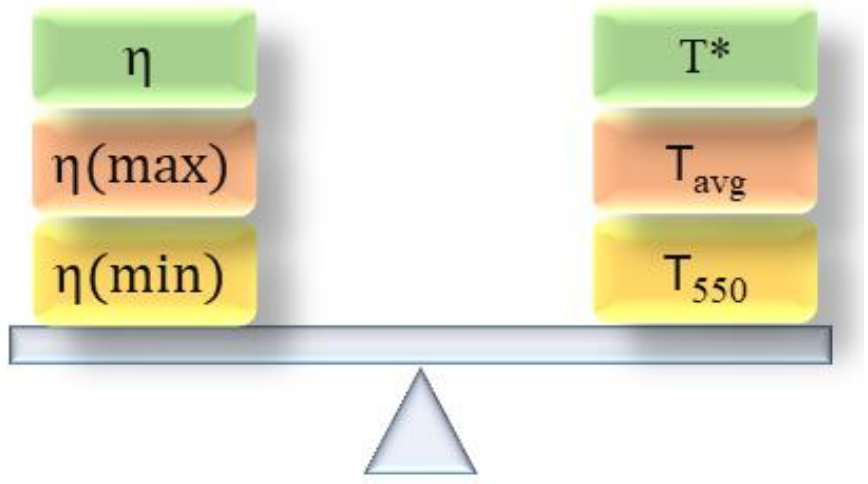

**Figure 1.** FoM balances the counter running parameters.

Due to the potential and clear disruptive technology the ST-SCs possess, researchers have been able to develop varieties of ST-SCs and depending on the device architecture, type of absorbing layer and other device layer materials, a wide range of transparency and efficiencies have been reported. Depending on the targeted application, the selection of the most suitable ST-SC is crucial and can be assessed by so-called figure-of-merit (FoM). An FoM is often defined as an appropriate quantitative measure that can rate the performance of the evaluated object and ascertain its comparative effectiveness in order to determine its relative utility for an application. To our knowledge, there is no such FoM defined for ST-SCs as of now.

In this work, we have proposed an FoM for ST-SCs taking into account various parameters. We took data for various device architecture of ST-SCs from literature and did a comparative study of all the parameters required to calculate the FoM of these ST-SCs. As depicted in Figure 1, an FoM helps in finding a trade-off between transparency and power conversion efficiency of ST-SCs considering all important parameters. Based on the quantitative measurement done one can rate the performance of the investigated device. To ensure the wide acceptability of the FoM for ST-SCs, we have mainly focused on major parameters that are easily accessible from most of the experimental measurements.

## TECHNICAL DISCUSSION ON PROPOSED FIGURE-OF-MERIT

Typically, in the broader context, the performance of a solar cell is quantified by its power conversion efficiency (PCE). However, while defining the quality of an ST-SC, the transparency of the device must also be taken into account. For making a good ST-SC, there has always been a demand for a minimal compromise between these two parameters. In the proposed FoM, we have emphasized equally the transmittance as well as the efficiency of the photovoltaic device.

Three transmittance terms are introduced which are material-specific and wavelength-dependent parameters. Since the transparency of an ST-SC largely depends on the active/absorbing layer material and may vary from material to material, therefore first transmittance term used is T* which is the estimated upper limit of transmittance of the active layer in the working wavelength regime and corresponding to the energy bandgap ($E_g$) of the active layer material. First, the wavelength below which absorption takes place is calculated as:

$$\lambda_o = \frac{\mathrm{hc}}{E_g} \tag{1}$$

where, h is Planck's constant, c is the speed of light in vacuum, $E_g$ is the energy bandgap corresponding to the active layer and $\lambda_o$ is the cut-off wavelength for absorption of light in the active layer material i.e., the light gets absorbed by the active layer material below this wavelength limit.

To set an upper limit for transmittance, we assume that above the cut-off wavelength, all the light will be transmitted, i.e., the transmittance is 100% and below the cut-off wavelength, 80% of the incident light is getting transmitted. The reason for choosing such high transmittance (80%) below the cut-off wavelength is because of the fact that for some applications, having a high transmittance is prioritised compared to achieving high-efficiency ST-SCs which can be achieved by significantly reducing the thickness of the active layer. There are works in which transmittance of ~60% (at 550nm) with an efficiency of 5.11% has already been demonstrated [11] and we believe that with optimized device architecture and proper choice of material, transparency close to 80% can be achieved. Similarly, having transmittance >80% below the cut-off will not help as then very few photons will be left to get absorbed resulting in negligible PCE. So, by knowing the value of $\lambda_o$, we can calculate the upper limit of transmittance (T*) as follows:

$$T^* = \frac{(\lambda_{max} - \lambda_o) * 1}{(\lambda_{max} - \lambda_{min})} + \frac{(\lambda_o - \lambda_{min}) * 0.8}{(\lambda_{max} - \lambda_{min})} \quad ; \ \lambda o < \lambda \max \qquad (2)$$

$$= 0.8 \quad ; \ \lambda o \geq \lambda \max$$

where $\lambda_{max}$ and $\lambda_{min}$ are the upper and lower wavelength limit of the regime of interest for which T* is calculated. Even though the ST-SCs are mostly known for their wide application in BIPV applications but there is another way to get more efficient output from these ST-SC by combining them with other solar cells resulting in a tandem solar cell. For example, a tandem solar cell consisting of a silicon cell overlaid by a PvSC could increase efficiencies of commercially mass-produced photovoltaics beyond the single-junction cell limit without adding substantial cost. Eventually, in this work, we have proposed two regimes under which FoM for ST-SCs is defined depending on the application for which they are being used, the first being for BIPV and the other is for tandem solar cells applications. Accordingly, we considered 375nm–700nm and 300nm–1100nm as the lower ($\lambda_{min}$) and upper ($\lambda_{max}$) wavelengths limits for BIPV and tandem applications, respectively. As can be noted from the limits, for BIPV, the wavelength range is taken considering only the visible region of the light. For tandem applications, generally, the bottom cell being used is a silicon solar cell having a bandgap of 1.12eV, which corresponds to the wavelength of about 1100nm and therefore the $\lambda_{max}$ is chosen accordingly. For tandem cells having different bottom cells other than silicon, the $\lambda_{max}$ value shall vary accordingly. Also, depending upon the thickness of different (mainly active) layers of the top cell, as discussed before, it may allow some transmittance even in the visible region which is beneficial for the bottom cell for higher charge generation. Taking this into account, the lower wavelength limit, $\lambda_{min}$ is kept at 300 nm as the most commonly used Si bottom cell absorbs light even in the UV regime.

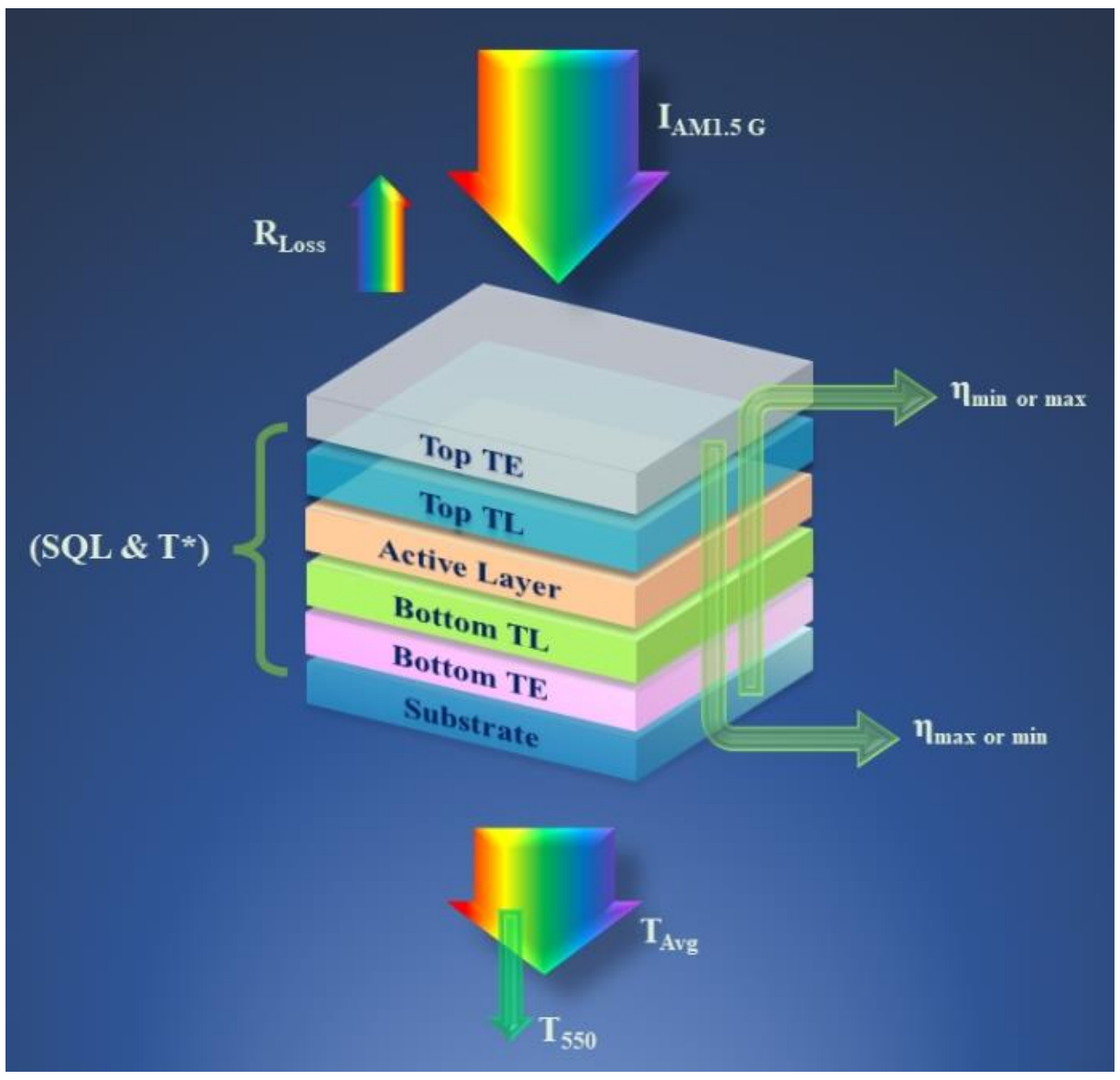

**Figure 2.** Schematic diagram of semi-transparent solar cell.

As, the wavelengths transmitted through the device decides the transparency of the device, in the transmitted portion of light what matters is the light that can be used for visibility in BIPV or improvement of PCE of the bottom cell in tandem configuration. Therefore, in deciding the transparency of an ST-SC the second transmittance term used is $T_{Avg}$ which is the average visible transmittance (AVT) of the device within the desired region depending upon application i.e. 375nm-700nm for BIPV and 300nm-1100nm for tandem application. Third and the last transmittance term used is $T_{550}$ which is the transmittance of the device corresponding to 550nm wavelength. Apart from T* and $T_{Avg}$ in which transmittance of the device is analysed under broad range spectrum equal importance is given to $T_{550}$ because the photopic response of the human eye is maximum near 550nm, which means better transmittance at 550nm can result in better visibility through ST-SCs.

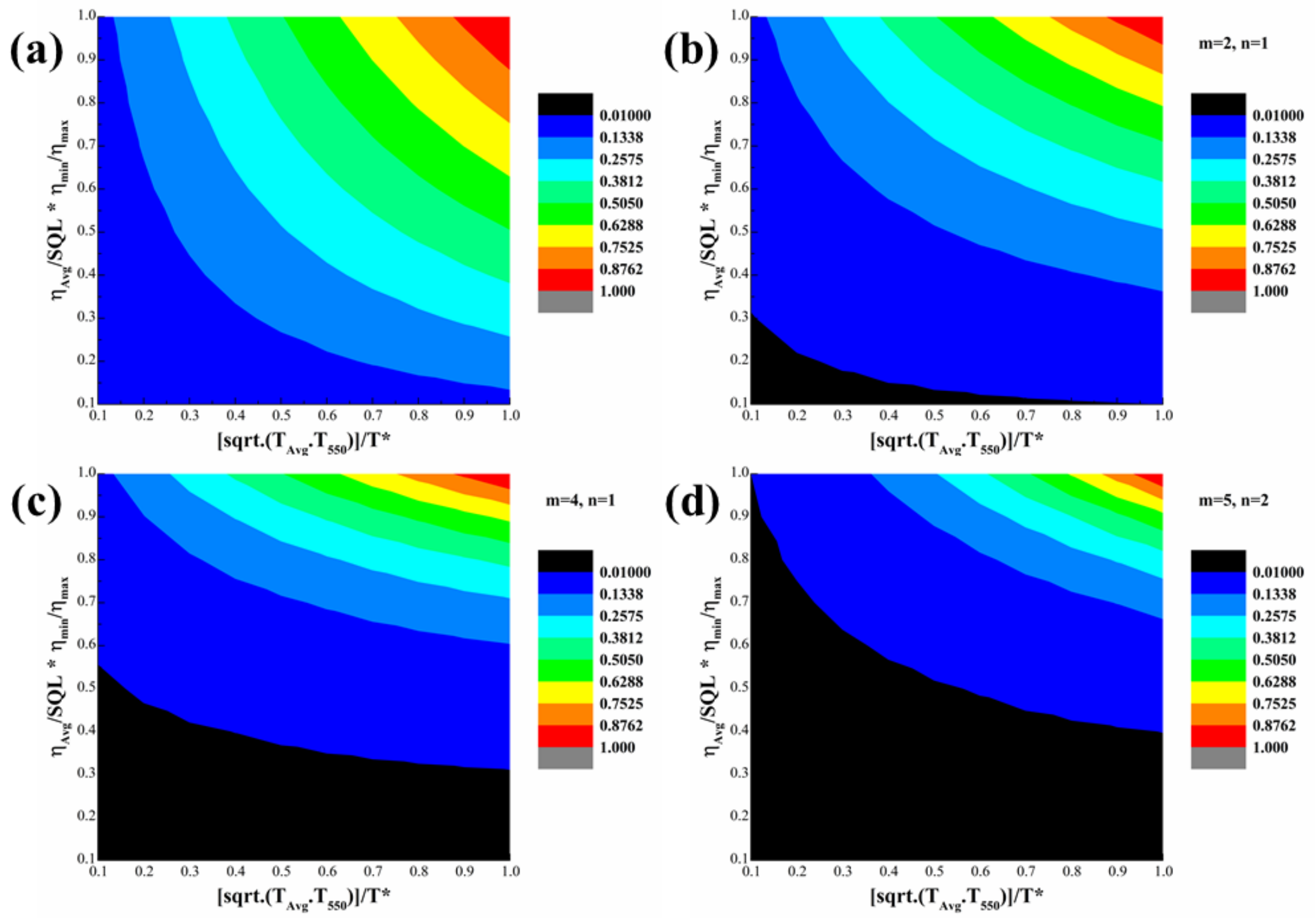

**Figure 3** Variation of figure of merit w.r.t. variation in possible values of efficiency and transmittance terms using different combinations of exponent (m and n) values for transmittance and efficiency terms (a) m=1, n=1, (b) m=2, n=1, (c) m=4, n=1, and (d) m=5, n=2.

Figure 2 shows a general device architecture of a semi-transparent solar cell. When light incidents on the semi-transparent solar cell irrespective of the direction of incident light (either from bottom or top electrode), depending on the substrate used, layers and their thicknesses, some portion of the light is reflected, some get transmitted and the rest of it is absorbed mainly by the active layer. For an ST-SC, there might be a difference in power conversion efficiency (PCE) while illuminating the device from different sides which is due to the asymmetric layers the light encounters while passing through the device from either side and the corresponding exciton generation. Accordingly, we define $\eta_{max}$ as the higher PCE while $\eta_{min}$ as the smaller value among the both while $\eta_{Avg}$ is the average PCE for the device and is the arithmetic mean of $\eta_{max}$ and $\eta_{min}$. The last term used in FoM is SQL which is the Shockley Queisser Limit for PCE [12].

As discussed earlier, the biggest challenge for defining an FoM for an ST-SC is to properly relate the transparency and efficiency of the investigated device as a photovoltaic device can have good transparency but low efficiency and vice-versa. Thus, our FoM is defined in such a way that for an ST-SC, the transparency and efficiency affect it equivalently. Therefore, the final FoM for the BIPV application is defined as:

$$FoM_{BIPV} = \left(\frac{\eta_{Avg}}{\text{SQL}} \times \frac{\eta_{min}}{\eta_{max}}\right) \times \left(\frac{\sqrt{[\text{T}_{Avg} \cdot \text{T}_{550}]}}{\text{T}*}\right) \quad (3)$$

On the other hand in the case of a tandem solar cell, the bifaciality factor $\left({}^{\eta_{min}}/_{\eta_{max}}\right)$ of the top cell does not play any role as only the dominant efficiency side is of interest. As only one value (higher one) of efficiency is considered, $\eta_{Avg}$ will be replaced by $\eta$. Also, here the transmittance value corresponding to 550 nm is not playing any significant role, so it has been removed from the expression and the final expression of FoM for tandem application is defined as:

$$FoM_{TDM} = \left(\frac{\eta}{\text{SQL}}\right) \times \left(\frac{\text{T}_{Avg}}{\text{T}*}\right) \quad (4)$$

It has to be noted that, in the above expressions for FoM, as discussed, we have given equal weightage for efficiency and transmittance. To validate it further, we tried generalizing the formula by taking exponents *m* and *n* for efficiency and transmittance terms respectively (equation (5).

$$\text{FoM}_{BIPV} = \left(\frac{\eta_{Avg}}{\text{SQL}} \times \frac{\eta_{min}}{\eta_{max}}\right)^{m} \times \left(\frac{\sqrt{[\text{T}_{Avg} \cdot \text{T}_{550}]}}{\text{T}*}\right)^{n} \quad (5a)$$

$$\text{FoM}_{TDM} = \left(\frac{\eta}{\text{SQL}}\right)^{m} \times \left(\frac{\text{T}_{Avg}}{\text{T}*}\right)^{n} \quad (5b)$$

In both the formulas, the maximum value of both the efficiency and transmittance terms can be 1 (as the efficiency and transmittance have been divided by their maximum limits). We made a contour plot representing FoM for different combinations of efficiency and transmittance term keeping the values of *m* and *n* equal to 1. As can be seen in figure 3, the contour is symmetrical about both the axes (efficiency and transmittance terms). On the other hand, when we take values of *m* and *n* other than 1, it leads to an asymmetric plot which means dependence on one of the terms is different from another which will be inappropriate as both the efficiency and transmittance are equally important while making a ST-SC. So, taking this into the consideration, the *m* and *n* values are kept at 1.

**Table I. T* corresponding to the bandgap ($E_g$) of active layer material for BIPV and tandem application**

| Eg (eV) | T* (Building Integrated) | T* (Tandem) |
|---|---|---|
| 0.4 | 0.8 | 0.8 |
| 0.7 | 0.8 | 0.8 |
| 1 | 0.8 | 0.8 |
| 1.3 | 0.8 | 0.84 |
| 1.6 | 0.8 | 0.88 |
| 1.9 | 0.83 | 0.91 |
| 2.2 | 0.88 | 0.93 |
| 2.5 | 0.93 | 0.95 |
| 2.8 | 0.96 | 0.96 |
| 3.1 | 0.98 | 0.98 |

Table I shows, an example of the values of T* for an ST-SC calculated using equation 2 for both BIPV and tandem application for different bandgap values of the active or absorbing layer. We also did an extensive literature survey for ST-SCs and tabulated them in Table II. Using the SQL and T* value corresponding to the bandgap of the active layer material from Table I and using all input parameters from various reported research articles, FoM is calculated in BIPV and tandem device configuration using equations 3 and 4. Some of the values in the table cells are missing due to the unavailability of corresponding data in the reference. In addition, we couldn't find corresponding parameters for the calculations of FoM for semi-transparent organic/polymer solar cells and therefore not included in the table [13–20]. Also, due to limited data points for $FoM_{TDM}$, most of the calculations and plots has been made for $FoM_{BIPV}$ only.

**Table II. Calculated FoM for different ST-SCs**

| Device architecture | $E_g$ (eV) | T* (BIPV) | T* (Tandem) | $T_{Avg}$ (375-700nm) (%) | $T_{Avg}$ (300-1100nm) (%) | $T_{550}$ (%) | η(min) (%) | η(max) (%) | η (%) | $FoM_{BIPV}$ | $FoM_{TDM}$ | Ref. |
|---|---|---|---|---|---|---|---|---|---|---|---|---|
| glass/ITO/PEDOT:PSS /CH3NH3PbI3−xClx / PCBM/BCP /ultrathin- Ag/MoO3, | 1.53 | 0.8 | 0.87 | 8.28 | - | 7.08 | 8.34 | 13.38 | 10.86 | 0.019 | - | [21] |
| glass/fto/tio2/perovskite/ spiro-OMeTAD /PEDOT:PSS /PDMS/PMMA/Graphene (150nm active layer) | 1.55 | 0.8 | 0.88 | 22.01 | - | 19.68 | 5.7 | 5.98 | 5.84 | 0.043 | - | [22] |
| glass/fto/tio2/perovskite /spiro-OMeTAD /PEDOT:PSS /PDMS/PMMA/Graphene (170nm active layer) | 1.55 | 0.8 | 0.88 | 19.55 | - | 19.02 | 6.62 | 7.22 | 6.92 | 0.045 | - | [22] |
| glass/fto/tio2/perovskite /spiro-OMeTAD/ PEDOT:PSS /PDMS/PMMA/Graphene | 1.55 | 0.8 | 0.88 | 17.12 | - | 12.03 | 7.95 | 8.72 | 8.34 | 0.04 | - | [22] |
| (200nm active layer) | | | | | | | | | | | | |
| glass/fto/tio2/perovskite /spiro-OMeTAD /PEDOT:PSS /PDMS/PMMA /Graphene (250nm active layer) | 1.55 | 0.8 | 0.88 | 11.4 | - | 6.12 | 9.15 | 10.1 | 9.63 | 0.027 | - | [22] |
| glass/fto/tio2/perovskite /spiro-OMeTAD /PEDOT:PSS /PDMS/PMMA /Graphene (290nm active layer) | 1.55 | 0.8 | 0.88 | 8.6 | - | 3.93 | 10.23 | 10.67 | 10.45 | 0.022 | - | [22] |
| glass/fto/tio2/perovskite /spiro-OMeTAD /PEDOT:PSS /PDMS/PMMA /Graphene (350nm active layer) | 1.55 | 0.8 | 0.88 | 7.16 | - | 2.18 | 11.65 | 12.02 | 11.84 | 0.017 | - | [22] |
| glass/FTO/compact TiOx/ mesoporous TiOx / PCBM:PMMA/ FA0.75Cs0.25Pb (I0.80Br0.20)3 /Spiro-OMeTAD/ITO/MgF2 | 1.7 | 0.8 | 0.89 | 5.61 | - | 1.59 | 13.4 | 14.6 | 14 | 0.014 | - | [23] |
| glass/Au/PEDOT:PSS/ CH3NH3PbI3-xClx/ PCBM/DMD (15nm gold) | 1.55 | 0.8 | 0.88 | 9.56 | - | 1.5 | 3.43 | 4.01 | 3.72 | 0.004 | - | [24] |
| glass/Au/PEDOT:PSS/ CH3NH3PbI3-xClx/ PCBM/DMD (20nm gold) | 1.55 | 0.8 | 0.88 | 8.92 | - | 1.2 | 3.95 | 6.45 | 5.2 | 0.004 | - | [24] |
| glass/Au/PEDOT:PSS/ CH3NH3PbI3-xClx/ PCBM/DMD (25nm gold) | 1.55 | 0.8 | 0.88 | 6.38 | - | 1.85 | 3.87 | 8.05 | 5.96 | 0.004 | - | [24] |
| glass/Au/PEDOT:PSS/ CH3NH3PbI3-xClx/ PCBM/DMD (30nm gold) | 1.55 | 0.8 | 0.88 | 5.37 | - | 1.06 | 3.63 | 8.6 | 6.11 | 0.002 | - | [24] |
| glass/FTO/mesoporous TiO2/Perovskite/ Spiro-OMeTAD/ITO | 1.57 | 0.8 | 0.88 | 3.48 | - | 0.95 | 11.7 | 15.6 | 13.65 | 0.007 | - | [25] |
| glass/FTO/ZnO/PCBM /Perovskite/Spiro-OMeTAD /In2O3:H | 1.55 | 0.8 | 0.88 | 10.86 | 38.6 | 3.66 | 9.5 | 14 | 11.75 | 0.019 | 0.15 | [26] |
| glass/ITO/PEDOT:PSS/Perovs kite /PCBM/AZO/ITO | 1.5 | 0.8 | 0.86 | 6.44 | - | 2.65 | 5.24 | 5.87 | 5.555 | 0.008 | - | [27] |
| glass/FTO/mesoporous TiO2 /Perovskite/CuSCN/ITO | 1.58 | 0.8 | 0.88 | 4.9 | - | 1.29 | 12.7 | 13.7 | 13.2 | 0.011 | - | [28] |
| glass/ITO.PEDOT:PSS/ Perovskite/PCBM/BCP/ Ag/V2O5 | 1.58 | 0.8 | 0.88 | 5.38 | - | 4.46 | 8.91 | 14.01 | 11.46 | 0.013 | - | [29] |
| ITO/PEDOT:PSS/perovskite /PCBM/BCP/IZO | 1.55 | 0.8 | 0.88 | 7.26 | - | 1.37 | 11.2 | 12.8 | 12 | 0.012 | - | [30] |
| NOA63/MoO3⁄Au /PEDOT:PSS /Perovskite/PCBM /MoO3⁄Au⁄Ag⁄MoO3⁄Alq3 | 1.54 | 0.8 | 0.86 | 11.85 | - | 9.48 | 5.83 | 6.9 | 6.36 | 0.021 | - | [31] |
| TETA-Graphene/ZnO /MAPbI3 /PTAA/PEDOT:PSS /TFSA-Graphene | 1.52 | 0.8 | 0.86 | 7.39 | - | 2.56 | 10.5 | 10.73 | 10.64 | 0.017 | - | [32] |
| ITO/PEDOT/CH3NH3PbI3 /PCBM/BCP/Ag/ | 2 | 0.85 | 0.92 | 15.62 | - | 16.14 | 8.7 | 11 | 9.85 | 0.043 | - | [33] |

| | | | | | | | | | | | |
|---|---|---|---|---|---|---|---|---|---|---|---|
| MoO3 (240nm active layer) | | | | | | | | | | | |
| ITO/PEDOT/CH3NH3PbI3 /PCBM/BCP/Ag/MoO3 (150nm active layer) | 2 | 0.85 | 0.92 | 19.21 | - | 26.06 | 5.4 | 7.3 | 6.35 | 0.037 | - | [33] |
| ITO/PTAA/CsPbI3 QDs/C60/BCP/Graphene | 1.8 | 0.80 | 0.90 | 56.15 | - | 59.89 | 4.95 | 5.11 | 5.03 | 0.105 | - | [11] |
| FTO/c-$TiO_2$/m-$TiO_2$/MAPbI3/Spiro/Cu(8nm) | 1.55 | 0.8 | 0.88 | 13.18 | 30.82 | 12.012 | - | 5.07 | | - | 0.05 | [34] |

It is interesting to note that while calculating $FoM_{BIPV}$ and $FoM_{TDM}$, for the same device, the value of the latter will always be greater than the former because- (i) the bifaciality factor is not there in the expression of $FoM_{TDM}$ which is always $\leq 1$, (ii) $T_{550}$ (which for most of the cases is below $T_{Avg}$) is only there in the expression of $FoM_{BIPV}$, and (iii) also the wavelength range for tandem application is more than that for BIPV application, so for the ST-SCs with bandgap higher than that of the bottom cell (generally silicon solar cell) the average transmittance in the range (300-1100) nm is going to be higher than that in the visible region.

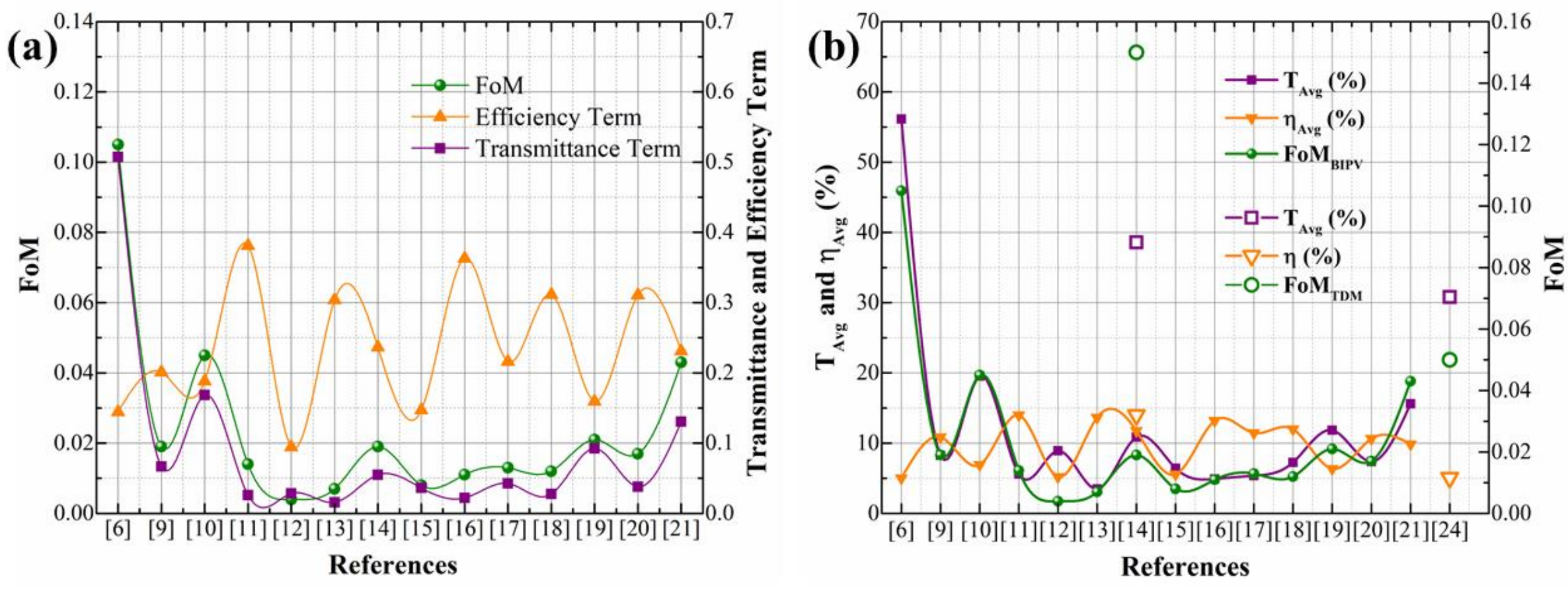

**Figure 4.** (a) Variation of $FoM_{BIPV}$ with transmittance term and efficiency term. (b) Variation of $FoM_{BIPV}$ and $FoM_{TDM}$ with average transmittance and efficiency.

In figure 4(a), we have shown the variation of $FoM_{BIPV}$ with transmittance term and efficiency term. It has been observed from figure 4(a) that the dependence of FoM on both the transmittance and efficiency term is identical (i.e., for two different devices with similar efficiency terms, the variation of FoM is similar to that of transmittance term and vice-versa). Using the same data values in the table above, we have also plotted the variation of $FoM_{BIPV}$ and $FoM_{TDM}$ with average device transmittance and device PCE (figure 4(b)) which are more easily relatable to the experimental results of an ST-SCs. Due to limited data availability in literature, $FoM_{TDM}$ has been calculated for only two references. As can be seen in the figure, for reference [11] and [13] where transmittance term is almost equal in both cases but there is some change in efficiency term and accordingly the FoM varies, from references [17] and [21] it is also observed that when efficiency term is almost equal but transmittance term is different than FoM value is varying according to transmittance term. This is what we aimed at the beginning that the FoM should have equal dependence on efficiency and transmittance. Similarly, we can also plot the dependence of $FoM_{TDM}$ with respect to transmittance and efficiency terms, but due to the unavailability of transmittance data in desired wavelength range, we haven't plotted the same. As mentioned at the beginning, to realize ST-SCs, the active layer material, its thickness and the type of transparent top electrodes greatly influences the FoM. Figure 5 shows the variation of FoM of different literature works (Table II) accordingly.

We have also considered some hypothetical scenarios to show the dependence of FoM on various factors. First of all, we considered the case where the efficiency of the device is the same irrespective of the illumination side (bifaciality factor = 1) and we studied the variation of $FoM_{BIPV}$ by varying the transmittance term in its possible range for different values of efficiency term. In figure 6(a), we can see that $FoM_{BIPV}$ is increasing linearly with an increase in transmittance term for different values of efficiency term. A similar plot can also be obtained for $FoM_{TDM}$ by varying the transmittance and efficiency terms. In another hypothetical case, we consider devices having efficiency equal to 30% of SQL and plot the variation of $FoM_{BIPV}$ w.r.t. transmittance term for different values of bifaciality factor. In this case as well, the variation is linear with respect to both transmittance as well as bifaciality factor (figure 6(b)).

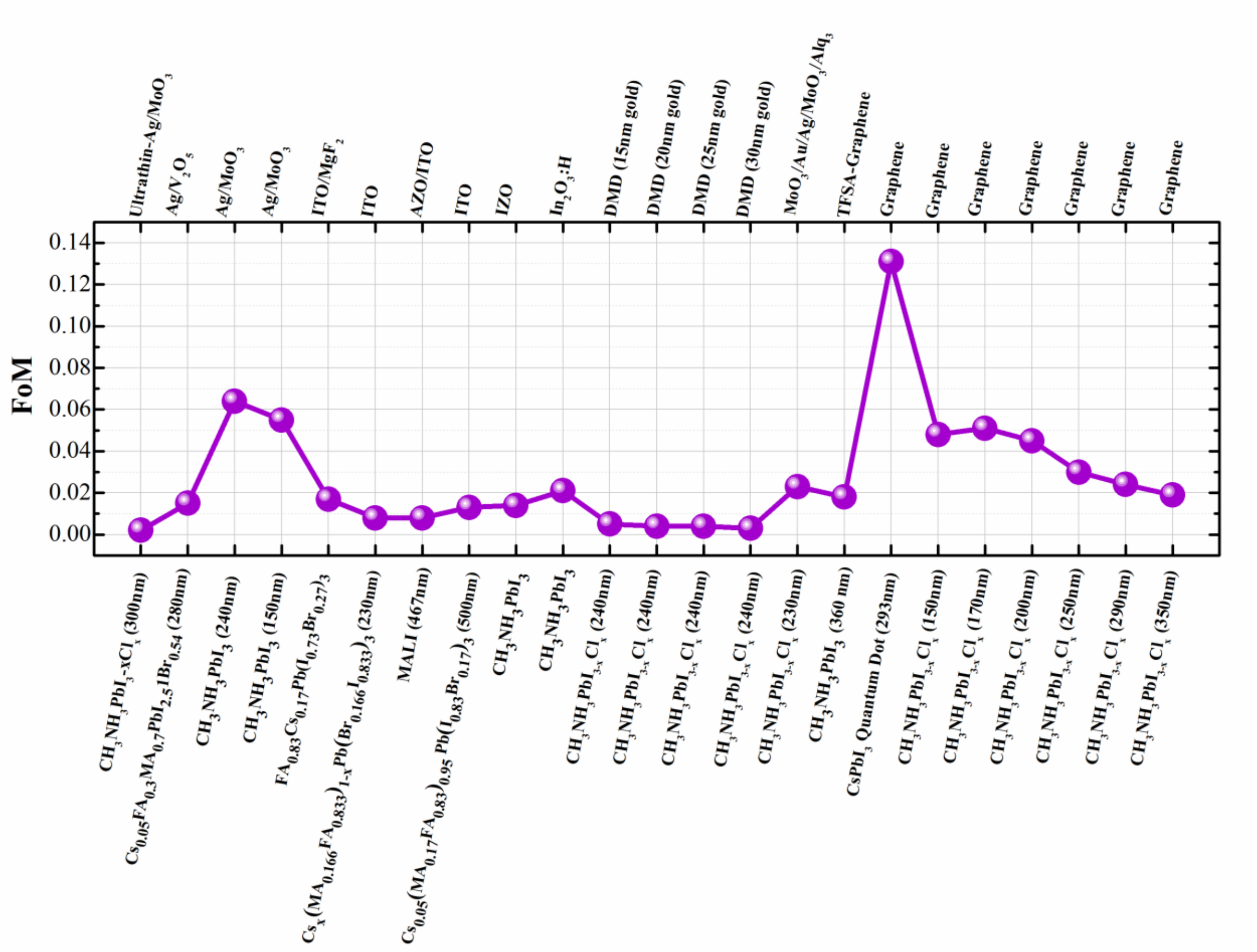

**Figure 5.** Variation of FoM with different active and transparent conducting layers.

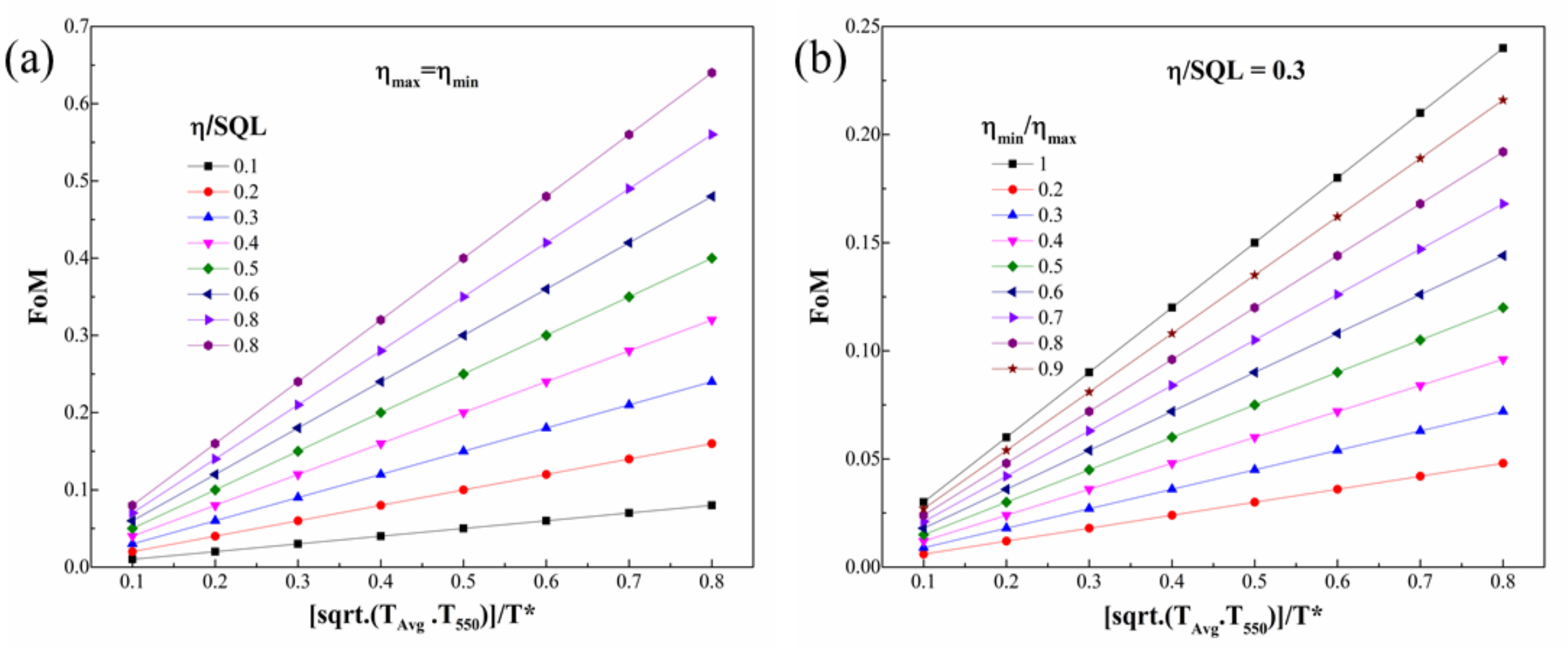

**Figure 6.** (a) Variation of FoM with transmittance term and efficiency term. (b) variation of FoM with transmittance term and bifaciality factor.

## CONCLUSION

In summary, we proposed, for the first time, an FoM to assess and compare different ST-SCs having different combinations of device efficiency and transparency. The FoM provide us the tool to rate the performance of the ST-SCs and ascertain their comparative effectiveness in order to determine its relative utility for applications such as BIPV and tandem cells. The proposed FoM has a different expression for BIPV and tandem applications and gives equal weightage to device efficiency and optical transparency. The dependence of the FoM on the exponents of the efficiency and transparency terms is also investigated and found to be most suitable with exponents values equal to unity for both competing terms. Among others, the proposed FoM takes into consideration the efficiency, bifaciality factor, average visible transmittance of the device in the region of interest, and transmittance at 550nm. All the considered parameters can be easily measured with some of them part of standard device characterization. Using these parameters, we have calculated $FoM_{BIPV}$ and $FoM_{BIPV}$ for data available in the literature and the same has been tabulated to show the performance of various ST-SCs having different photoactive layers of different thicknesses and incorporating different transparent top electrodes. With the introduction of the FoM, we hope that it will contribute to guiding researchers' commitment to developing, evaluating, and assessing the continuing progress in the field of ST-SCs.

## ACKNOWLEDGEMENT

DSG acknowledge funding support from Department of Science and Technology (DST) of India via grants DST/NM/NT/2018/146, ECR/2018/002132 and SB/S2/RJN-048/2016. AK and SR acknowledge Prime Minister Research Fellowship (PMRF-192002-1086 and PMRF-192002-2025 respectively).